\documentclass[aps,prb,reprint,superscriptaddress,floatfix]{revtex4-2}

\usepackage{amsmath,amssymb,amsfonts}
\usepackage{graphicx}
\usepackage{dcolumn}
\usepackage{bm}
\usepackage{hyperref}

\hypersetup{
    colorlinks=true,
    linkcolor=blue,
    citecolor=blue,
    urlcolor=blue
}

\begin{document}

\title{Experimental Evidence for Type-I Superconductivity in $\mathrm{OsB_2}$}

\author{Yogesh Singh}
\email{yogesh@iisermohali.ac.in}
\affiliation{Department of Physical Sciences, Indian Institute of Science Education and Research (IISER) Mohali, Knowledge City, Sector 81, Mohali 140306, India.}

\date{\today}

\begin{abstract}
Since the discovery of superconductivity in $\mathrm{MgB_2}$, metal diborides have attracted significant attention. Among these, osmium diboride ($\mathrm{OsB_2}$), crystallizing in an orthorhombic structure (space group $Pmmn$) with puckered boron layers, has been classified as a low-$\kappa$ Type-II superconductor ($T_c \approx 2.1~\mathrm{K}$). Here, we report experimental evidence establishing that $\mathrm{OsB_2}$ is in fact a rare nonelemental Type-I superconductor. Our identification is supported by three primary experimental hallmarks: (i) a pronounced Differential Paramagnetic Effect (DPE) peak in AC susceptibility ($\chi_{ac}$) measurements under static DC magnetic fields, signaling the intermediate state; (ii) a field-induced crossover to a divergent first-order phase transition in specific heat ($C/T$); and (iii) pronounced magnetic hysteresis in field-ramped transitions indicative of supercooling. 
\end{abstract}

\maketitle

\section{Introduction}

The discovery of multi-gap superconductivity in $\mathrm{MgB_2}$ ~\cite{MgB2a, MgB2b} has stimulated renewed interest in research on transition-metal diborides~\cite{Singh2007, Singh2010, Suh2007, Singh2018, Datta2020, Mehta2026, Gupta2026}. Osmium and Ruthenium diboride crystallize in an orthorhombic structure with space group $Pmmn$ (No.~59), consisting of corrugated hexagonal boron layers stacked along the $c$-axis between osmium planes~\cite{Singh2007, Singh2010}. Early studies on polycrystalline and single-crystal samples established bulk superconductivity below $T_c \approx 2.1~\mathrm{K}$ for OsB$_2$~\cite{Singh2007, Singh2010} and $T_c \approx 1.6~\mathrm{K}$ for RuB$_2$~\cite{Singh2007, Singh2018}. $\mathrm{OsB_2}$ was historically assigned as a moderate-coupling, low-$\kappa$ Type-II, electron-phonon-mediated  superconductor with a Ginzburg-Landau parameter $\kappa \approx 1\text{--}2$~\cite{Singh2007, Singh2010, Suh2007}.

Despite this Type-II classification, several experimental anomalies—including a reduced specific heat jump at $T_c$ ($\Delta C / \gamma T_c \approx 0.94\text{--}1.32$), Type-I like magnetization curves $M(H)$, and an unusual concave curvature in the $H_c\text{--}T$ phase boundary—prompted a re-examination of the nature of its superconducting state~\cite{Singh2007, Singh2010, Suh2007}. In particular, theoretical electronic structure calculations using density functional theory (DFT) and Shubnikov-de Haas oscillations revealed that $\mathrm{OsB_2}$ possesses multi-sheet Fermi surfaces with remarkably high Fermi velocities ($v_F \approx 3.71 \times 10^7\text{ to }9.1 \times 10^7~\mathrm{cm/s}$) and almost free electron-like effective masses~\cite{Bekaert2015, Bekaert2016, Singh2010}. Because the BCS coherence length is directly proportional to the Fermi velocity,
\begin{equation}
\xi_0 = \frac{\hbar v_F}{\pi \Delta(0)},
\end{equation}
the high Fermi velocities result in large coherence lengths ($\xi_0 \approx 133\text{--}315~\mathrm{nm}$), suppressing the Ginzburg-Landau parameter $\kappa = \lambda/\xi$ to $0.10\text{--}0.24$~\cite{Bekaert2015, Bekaert2016}. These theoretical estimates place $\mathrm{OsB_2}$ deep inside the Type-I regime ($\kappa < 1/\sqrt{2} \approx 0.707$).  

Recently both OsB$_2$ and RuB$_2$ have been reported to be topological $Z_2$ metals with topological surface states~\cite{Mehta2026, Gupta2026}. Additionally, time reversal breaking has been observed on entering the superconducting state in RuB$_2$~\cite{Gupta2026}.  These recent discoveries have brought the question about the Type of superconductivity in these diborides in the limelight again.

In this paper, we present experimental evidence establishing $\mathrm{OsB_2}$ as a Type-I superconductor, joining a small group of binary intermetallics such as $\mathrm{PdTe_2}$ and $\mathrm{YbSb_2}$~\cite{Leng2019, Gao2026}.

\section{Experimental Methods}

High-purity single crystals of $\mathrm{OsB_2}$ were synthesized via arc-melting and solution growth using a Cu-B flux as reported previously~\cite{Singh2007, Singh2010}. The obtained crystals are thin hexagonal plates with width $\sim 3$~mm and thickness $\approx 0.4$~mm.  Phase purity of the crystals was confirmed via powder X-ray diffraction (XRD). Dynamic AC magnetic susceptibility ($\chi_{ac}$) was measured using the ac $\chi$ option in a physical property measurement system (PPMS-ECII). An ac field of 0.1~Oe at a frequency of 111~Hz was used to measure ac chi at various static dc fields applied parallel to the hexagonal plates. Specific heat at different temperatures and magnetic fields $C(T,H)$ was recorded down to $1.7~\mathrm{K}$ using a thermal relaxation technique in the PPMS.  Data on two crystals (Xtal1 and Xtal2) from different growth batches have been used for heat capacity. The Xtal1 had a mass of $16$~mg whereas Xtal2 had a mass of $4$~mg.  AC $\chi$ was measured on Xtal1. 

\section{Results and Discussion}

\subsection{Differential Paramagnetic Effect in AC Susceptibility}

A defining signature of Type-I superconductivity is the formation of an intermediate state, where the material phase-separates into alternating normal and superconducting domains for applied fields $H_a < H_c$, where $H_c$ is the critical field, and $H_a$ is the applied field. Within this field range, the differential susceptibility $\partial M / \partial H_a$ becomes positive.  This phenomenon is called the differential paramagnetic effect and has been observed for example in several elemental Type-I superconductors~\cite{Hein1961}.

\begin{figure}[t]
  \centering
 \includegraphics[width=.9\linewidth]{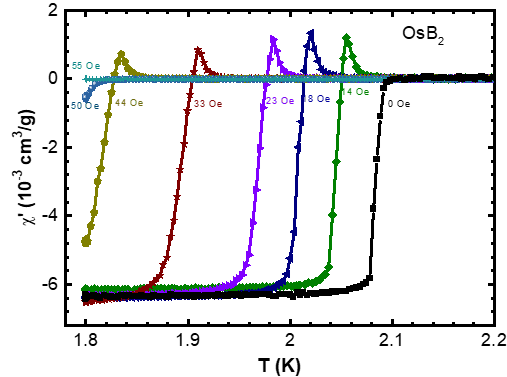}
  \caption{\label{fig:dpe}\textbf{Differential Paramagnetic Effect (DPE).} Real component of the AC susceptibility $\chi'(T)$ measured at $111~\mathrm{Hz}$ under fixed DC magnetic fields ($0 \le H_{dc} \le 100~\mathrm{Oe}$). In zero field, a sharp diamagnetic screening drop marks $T_c \approx 2.1~\mathrm{K}$. Under applied DC fields, a distinct positive DPE peak appears immediately prior to the normal-state transition, reflecting the positive slope $\partial M / \partial H_a > 0$ characteristic of domain entry in the Type-I intermediate state.}
\end{figure}

Figure~\ref{fig:dpe} shows the temperature dependence of the real component of AC susceptibility as a function of temperature $\chi'(T)$ under various static DC magnetic fields.  To minimize demagnetization effects the crystal was oriented so that the dc magnetic field was in the ab-plane (parallel to the hexagonal crystal surface).  As shown in Fig.~\ref{fig:dpe}, while the zero-field measurements display a sharp diamagnetic drop at $T_c \approx 2.1$~K, applying a modest DC magnetic field creates a prominent positive peak in $\chi'(T)$ just below the critical temperature. This positive Differential Paramagnetic Effect (DPE) peak is a signature of Type-I superconductors entering the intermediate state.

\subsection{Field-Induced First-Order Phase Transition}

\begin{figure*}[t]
  \centering
  \includegraphics[width=\linewidth]{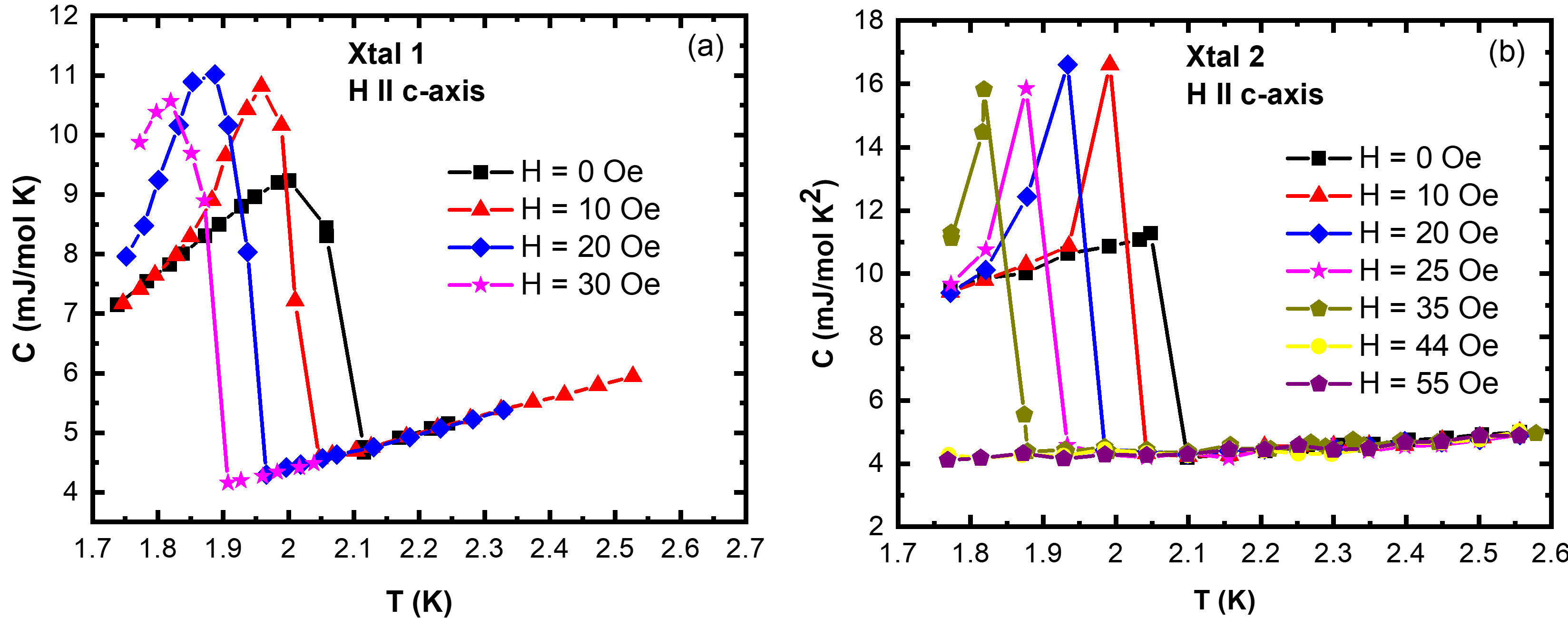}
\caption{\label{fig:specific_heat}\textbf{First-order superconducting transition in magnetic fields.} Temperature dependence of $C(T)$ measured on two different crystals under various magnetic fields. At $H=0$, the transition is a standard second-order jump with $\Delta C / \gamma T_c \approx 1.2$. Upon applying a magnetic field, the jump transforms into a sharp, divergent peak exceeding the zero-field amplitude, giving clear thermodynamic evidence for a field-induced first-order phase transition and latent heat.}
\end{figure*}

While zero-field superconducting transitions are second-order, Type-I superconductors in a finite magnetic field undergo a first-order phase transition accompanied by latent heat.
We find this signature as well for OsB$_2$. Figure~\ref{fig:specific_heat} displays the low-temperature specific heat $C(T)$ measured at various applied magnetic fields for two crystals from different growth batches.  The magnetic field was applied parallel to the $c$-axis in these measurements.

In zero field, $\mathrm{OsB_2}$ displays a step-like anomaly at $T_c \approx 2.1~\mathrm{K}$ consistent with previous reports~\cite{Singh2007,Singh2010}. When a magnetic field is applied, the transition sharpens into a sharp peak where $C$ dramatically exceeds its zero-field value. This behaviour at the normal-superconductor transition confirms the latent heat absorption characteristic of a field-induced first-order phase transition.\\

\subsection{Magnetic Hysteresis and Supercooling}
\begin{figure*}[t]
  \centering
   \includegraphics[width=\linewidth]{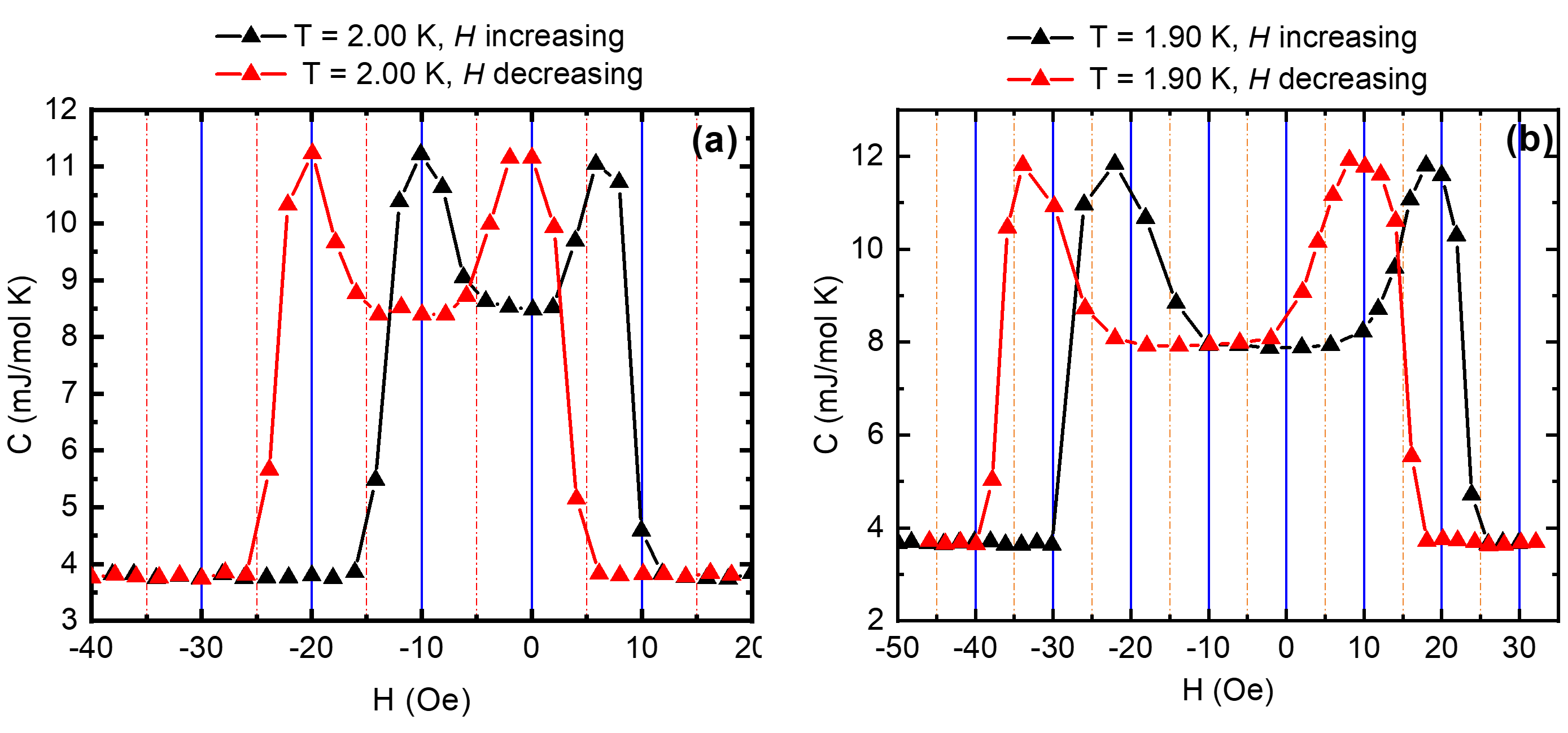}\caption{\label{fig:hysteresis}\textbf{Magnetic hysteresis and supercooling.} Specific heat $C(H)$ measured during field-increasing (black symbols) and field-decreasing (red symbols) sweeps at (a) $T = 2~\mathrm{K}$ and (b) $T = 1.9~\mathrm{K}$. A clear hysteresis is observed in the normal-superconducting phase boundary indicating supercooling.}
\end{figure*}

Type-I superconductors are expected to exhibit supercooling in a magnetic field, where the normal state persists metastably below the critical field $H_c$ as the field is ramped downward.  This can lead to hysteresis in the transition temperature as the magnetic field is increased or decreased through the normal-superconducting transition. Figure~\ref{fig:hysteresis} shows field-ramped heat capacity transition cycles measured at two different temperatures below $T_c$.  The hysteresis observed between field-up and field-down sweeps (Fig.~\ref{fig:hysteresis}) confirms the presence of supercooling during the superconducting-to-normal phase boundary traversal.

\section{Conclusion}

In summary, AC susceptibility, specific heat in fields, and field-ramping measurements across the normal-superconducting transition provide strong, unified evidence that $\mathrm{OsB_2}$ is a Type-I superconductor. The Differential Paramagnetic Effect (DPE), divergent specific heat anomaly in field, and field-ramping hysteresis confirm an intermediate state and first-order phase transition. These experimental observations are similar to recent reports on Type-I superconductivity in $\mathrm{PdTe_2}$ and $\mathrm{YbSb_2}$~\cite{Leng2019, Gao2026} and strongly suggest that $\mathrm{OsB_2}$ represents a rare example of a binary Type-I superconductor.

\begin{acknowledgments}
We thank the X-ray, the SEM, and the SQUID magnetometer facilities at IISER Mohali. We acknowledge DST, India, for support through grant CRG/2022/000015.
\end{acknowledgments}

\end{document}